\documentclass[aps,prl,twocolumn,amsmath,amssymb,showpacs]{revtex4}
\usepackage{graphicx}
\usepackage{Jonasmacros}
\usepackage{bm}

\begin{document}
\preprint{\today}
\title{Analytical Current Formula for Strongly Coupled Quantum Dot with Arbitrary Electron-Electron Interaction}
\author{J. Fransson}
\affiliation{Department of Materials Science and Engineering, Royal Institute of Technology (KTH), SE-100 44\ \ Stockholm, Sweden}
\affiliation{Physics Department, Uppsala University, Box 530, SE-751 21\ \ Uppsala, Sweden}

\begin{abstract}
An analytical expression for the current through a single level quantum dot for arbitrary strength of the on-site electron-electron interaction is derived beyond standard mean-field theory. By describing the localised states in terms of many-body operators, the employed diagrammatic technique for strong coupling enables inclusion of electron correlation effects into the description of the local dynamics, which provides transport properties that are consistent with recent experimental data.
\end{abstract}
\pacs{72.10.Bg, 73.63.Kv, 73.23.-b, 71.27.+a}
\maketitle

Electron transport through few electron islands is strongly influenced by electron correlations between its atomic like states. A variety of correlation effects, including Kondo effect \cite{kondo} and resonant current peaks \cite{resonant}, have been observed. For low conductance of the tunnel junctions one can employ "orthodox theory" \cite{orthodox}, which treats the tunnelling in lowest order perturbation theory (golden rule). In the strong coupling regime, however, this theory collapse since then the transport is not dominated by sequential tunnelling. While effects of coherent tunnelling have been extensively studied \cite{coherent}, the question of a general description which includes correlation effects and is valid both for strong and weak coupling regimes remains open. In this Letter, an analytical formula for the current through a single level quantum dot (QD) is derived for arbitrary on-site Coulomb repulsion, beyond standard mean-field theory (e.g. beyond self-consistent Hartree-Fock, or Hubbard I, approximation (HIA) \cite{hubbard,hewson1966,varma1976}). This formula  include local correlation effects which provide the well known renormalisation of the localised level \cite{barabanov1974,varma1976,renormalisation}. 

The present derivation concerns single level QDs, here described by the Anderson model \cite{anderson1961}, e.g.
\begin{eqnarray}
\Hamil&=&\sum_{k\sigma\in L,R}\leade{k}\cdagger{k}\c{k}
	+\sum_{p=0,\sigma,2}E_p\h{p}{}
\nonumber\\&&
	+\sum_{k\sigma\in L,R}
		[v_{k\sigma}\cdagger{k}(\X{0\sigma}{}+\eta_\sigma\X{\bar\sigma2}{})+H.c.],
\label{eq-model}
\end{eqnarray}
where $\cdagger{k}\ (\c{k})$ creates (annihilates) an electron in the left ($L$) or right ($R$) lead at the energy $\leade{k}$ and spin $\sigma=\up,\down$, whereas $\X{pq}{}=\ket{p}\bra{q}$, $\ket{p},\ket{q}=\ket{0},\ket{\sigma},\ket{2}\equiv\ket{\up\down}$, is a Hubbard operator \cite{hubbard}, where $\h{p}{}\equiv\X{pp}{}$ is the population operator for the state at the energy $E_p$. Here, the local state energies are $E_0=0$, $E_\sigma=\dote{\sigma}$, and $E_2=\dote{\up}+\dote{\down}+U$, where $\dote{\sigma}$ is the localised single electron energy, whereas $U$ is the on-site Coulomb repulsion. The last term in Eq. (\ref{eq-model}) models the tunnelling, with rate $v_{k\sigma}$, between the leads and QD. The factor $\eta_\sigma\equiv\bra{\bar\sigma}\d{\sigma}\ket{2}$ ($\eta_{\up,\down}=\pm1$) accounts for the selection rule between the singly and doubly occupied states ($\bra{0}\d{\sigma}\ket{\sigma}=1$), and $\bar\sigma$ is the opposite spin of $\sigma$.

Using the description of the Anderson model in terms of Hubbard operators is motivated because this enables the use of a diagrammatic technique which \emph{i)} provides an expansion of the QD GF in the strong coupling regime \cite{sandalov2003}, \emph{ii)} yields the logarithmic renormalisation of the localised level \cite{franssonPRL2002}, and \emph{iii)} can be employed for a direct transformation of the on-site properties into corresponding non-equilibrium quantities \cite{franssonPRB2004,franssonPRL2005}

The stationary current through the system can be calculated from the general expression Eq. (19) in Ref. \cite{jauho1994}, with the matrix $\bfGamma^{L/R}$ (given later) describing the coupling between the left/right lead and the QD, and the retarded/advanced/lesser matrix GF $\bfG^{r/a/<}$ for the local QD properties.

The components of the GF are defined by \cite{franssonPRL2002,sandalov2003,franssonPRB2004,franssonPRL2005}
$G_{a\bar{b}}(t,t')=\onlinenoneqgr{\X{a}{}(t)}{\X{\bar{b}}{}(t')}$, where $a,b$ denote Fermi like (single electron) transitions whereas $\bar{a},\bar{b}$ denote the conjugate transitions. The action $S=\exp{(-i\kbint\Hamil'(t)dt)}$, $1/\beta=k_BT$, \cite{franssonPRL2002,sandalov2003,franssonPRB2004,franssonPRL2005,kadanoff} contains time dependent disturbance source fields $U_\xi(t)$, used for the generation of a diagrammatic expansion of the GF in terms of functional derivatives. Here, $\Hamil'(t)=U_0(t)\h{0}{}+\sum_{\sigma\sigma'}U_{\sigma\sigma'}(t)\X{\sigma\sigma'}{}+U_2(t)\h{2}{}$. Physical quantities are accquired from the GF as $U_\xi(t)\rightarrow0$.

In the present system, there are four possible single electron transitions, e.g. $\X{0\sigma}{}$ and $\X{\bar\sigma2}{}$, which yields a $4\times4$ matrix equation for the GF. Following the derivation in Refs. \cite{franssonPRB2004,franssonPRL2002}, one finds that the general structure of the equation of motion for the matrix GF is given by
\begin{eqnarray}
\lefteqn{
\biggl(i\ddt I-\bfDelta^0-\bfU(t)\biggr)\bfG(t,t')=\delta(t-t')\bfP(t)
}
\nonumber\\&&
	+[\bfP(t^+)+\bfR(t^+)]\kbint\bfV(t,t'')\bfG(t'',t')dt'',
\label{eq-eqm}
\end{eqnarray}
where $I$ is the identity, $\bfDelta^0=\mbox{diag}\{\Delta_{\up0}^0,\Delta_{\down0}^0,\Delta_{2\down}^0,\Delta_{2\up}^0\}$ (diagonal matrix) contains the bare transition energies $\Delta_{\sigma0}^0=E_\sigma-E_0$ and $\Delta_{2\bar\sigma}^0=E_2-E_{\bar\sigma}$. The interaction between the leads and the QD is given by
\[
\bfV(t,t')=\left(\begin{array}{cc}
	\bfV(t,t') & \bfV'(t,t') \\ \bfV'(t,t') & \bfV(t,t')
		\end{array}\right),
\]
where $\bfV=\mbox{diag}\{V_\up,V_\down\}$ and $\bfV'=\sigma_z\bfV$ ($\sigma_z$ is the $z$-component of the Pauli spin vector), with $V_\sigma(t,t')=\sum_{k\in L,R}|v_{k\sigma}|^2g_{k\sigma}(t,t')$. Here, $g_{k\sigma}(t,t')$ is the GF for the conduction electrons in the leads, satisfying the equation $(i\ddtinline-\leade{k})g_{k\sigma}(t,t')=\delta(t-t')$.

In Eq. (\ref{eq-eqm}) the \emph{end-factor} $\bfP(t)$, which arise due to the non-commutative algebra of Hubbard operators, carries the spectral weights of the components of the matrix GF. In the considered system $\bfP=\mbox{diag}\{\bfP_1,\bfP_2\}$, where $\bfP_n,\ n=1,2$, themselves are $2\times2$ matrices given by $P_1=\{P_{0\sigma\sigma'0}\}_{\sigma\sigma'}\equiv\{\occu{\anticom{\X{0\sigma}{}}{\X{\sigma'0}{}}}\}_{\sigma\sigma'}$ and $P_2=\{P_{\bar\sigma22\bar\sigma'}\}_{\sigma\sigma'}\equiv\{\occu{\anticom{\X{\bar\sigma2}{}}{\X{2\bar\sigma'}{}}}\}_{\sigma\sigma'}$ (subscript $U$ signifying the source fields dependence of the averages). In the present case $P_{0\sigma\bar\sigma0},\ P_{\bar\sigma22\sigma}=0$. The non-commutativity of the Hubbard operators also gives rise to the operator $\bfR(t)$ \cite{sandalov2003,franssonPRB2004,franssonPRL2002}, here given by $\bfR=\mbox{diag}\{\bfR_1,\bfR_2\}$, where $\bfR_1(t)=\{R_{0\sigma\sigma'0}(t)\}_{\sigma\sigma'}\equiv\{\delta_{\sigma\sigma'}\delta/\delta U_0(t)+\delta/\delta U_{\sigma'\sigma}(t)\}_{\sigma\sigma'}$ and $\bfR_2(t)=\{R_{\bar\sigma22\bar\sigma'}(t)\}_{\sigma\sigma'}\equiv\{\delta_{\bar\sigma\bar\sigma'}\delta/\delta U_2(t)+\delta/\delta U_{\bar\sigma\bar\sigma'}(t)\}_{\sigma\sigma'}$
that is, functional differentiation operators which operate on the GF $\bfG$ with respect to the auxiliary source fields $U_\xi(t)$. The matrix $\bfU=\mbox{diag}\{\bfU_1,\bfU_2\}$, where $\bfU_1(t)=\{U_{\sigma\sigma'}(t)-\delta_{\sigma\sigma'}U_0(t)\}_{\sigma\sigma'}$ and $\bfU_2(t)=\{\delta_{\bar\sigma\bar\sigma'}U_2(t)-U_{\bar\sigma'\bar\sigma}(t)\}_{\sigma\sigma'}$.

The structure of the equation of motion, Eq. (\ref{eq-eqm}), suggests that $\bfG(t,t')=\bfD(t,t')\bfP(t')$, where $\bfD(t,t')$ is the \emph{locator} \cite{sandalov2003} carrying the local on-site properties of the GF, e.g. the position and width of the localised level. Here, fluctuations of the spectral weights are neglected, hence, $\bfP$ is treated as a constant with respect to the source fields $U_\xi(t)$. Therefore, varying the GF with respect to the source fields generates $\delta\bfG=(\delta\bfD)\bfP$, and since the locator satisfies the matrix property $\bfD\bfD^{-1}=I=\bfD^{-1}\bfD$ one finds that $\delta\bfD=-\bfD(\delta\bfD^{-1})\bfD$, hence $\delta\bfG=-\bfD(\delta\bfD^{-1})\bfG$. Thus, it is necessary to study the locator and its inverse.

In general, the equation of motion for the locator appear very similar to Eq. (\ref{eq-eqm}), that is, replacing $\bfG$ by $\bfD$ and removing $\bfP$ in the first term on the right hand side of Eq. (\ref{eq-eqm}). In addition, the structure of the resulting equation of motion for $\bfD$ suggests that the inverted locator $\bfD^{-1}$ can be written
\[
\bfD^{-1}(t,t')=\bfd^{-1}(t,t')-\bfS(t,t'),
\]
where the \emph{self operator} $\bfS(t,t')$ \cite{sandalov2003} is identified by
\begin{eqnarray*}
\bfS(t,t')=\{[\bfP(t^+)+\bfR(t^+)]
	\kbint\bfV(t,t_1)\bfD(t_1,t_2)\}
\nonumber\\
	\times\bfD^{-1}(t_2,t')dt_2dt_1,
\end{eqnarray*}
whereas the bare locator $\bfd(t,t')$ satisfies the equation $[i\ddtinline I-\bfDelta^0-\bfU(t)]\bfd(t,t')=\delta(t-t')I$. For the present derivation, the dressed inverted locator in $\delta\bfG=-\bfD(\delta\bfD^{-1})\bfG$ is replaced by its corresponding bare quantity, e.g. $\delta\bfG=-\bfD(\delta\bfd^{-1})\bfG$, meaning that the diagrammatic expansion terminate after the first functional differentiation, the so-called \emph{loop correction} \cite{franssonPRL2002,franssonPRB2004,sandalov2003}. This will then generate the renormalisation of the localised level, as is shown below. However, continuing the functional differentiation to higher orders generate higher order diagrams that account for additional many-body correlation effects \cite{franssonPRB2004,sandalov2003}, for instance contributions from the Kondo effect.

Due to the non-commutativity of the matrices in Eq. (\ref{eq-eqm}) it is now preferable to study the scalar equations. The interaction matrix $\bfV$ goes unaffected by the functional derivatives, since here the source fields are only applied to the localised QD states. Noticing that, for instance, $R_{0\sigma\sigma'0}(t^+)\bfd^{-1}(t,t')=-\delta(t-t')R_{0\sigma\sigma'0}(t^+)\bfU(t)$, one finds that the only non-vanishing contributions are
\begin{eqnarray*}
\left.\begin{array}{c}
R_{0\sigma\bar\sigma0}(t^+) \\
R_{\bar\sigma22\sigma}(t^+)
\end{array}\right\}G_{a\bar{b}}(t'',t)=
	i[D_{a\bar\sigma0}(t'',t^+)G_{0\sigma\bar{b}}(t^+,t')
\nonumber\\
	+D_{a2\sigma}(t'',t^+)G_{\bar\sigma2\bar{b}}(t^+,t')].
\end{eqnarray*}
The Fourier transformed components $G_{0\sigma\sigma0},\ G_{\bar\sigma2\sigma0}$ of Eq. (\ref{eq-eqm}), hence, reduces to
\begin{eqnarray*}
&&
(i\omega-\Delta_{\sigma0}-P_{0\sigma\sigma0}V_\sigma)
	G_{0\sigma\sigma0}(i\omega)=P_{0\sigma\sigma0}
\nonumber\\&&\hspace{2cm}
	+[\eta_\sigma P_{0\sigma\sigma0}V_\sigma
		+\eta_{\bar\sigma}\delta\Delta_{2\bar\sigma}]
	G_{\bar\sigma2\sigma0}(i\omega),
\\&&
	G_{\bar\sigma2\sigma0}(i\omega)=
	\frac{\eta_\sigma P_{\bar\sigma22\bar\sigma}V_\sigma
		+\eta_{\bar\sigma}\delta\Delta_{\sigma0}}
	{i\omega-\Delta_{2\bar\sigma}-P_{\bar\sigma22\bar\sigma}V_\sigma}
	G_{0\sigma\sigma0}(i\omega),
\end{eqnarray*}
(as $U_\xi(t)\rightarrow0$) and analogously for the components $G_{\bar\sigma22\bar\sigma},\ G_{\bar\sigma2\sigma0}$, with $\Delta_{\bar{a}}=\Delta_{\bar{a}}^0+\delta\Delta_{\bar{a}}$ and
\begin{subequations}
\label{eq-loop}
\begin{eqnarray}
\delta\Delta_{\sigma0}=\frac{1}{2\pi}\sum_{\alpha=L,R}
	\sum_{k\in\alpha}|v_{k\bar\sigma}|^2
	\int\frac{f_\alpha(\dote{k\bar\sigma})-f(\omega)}
		{\dote{k\bar\sigma}-\omega}
\nonumber\\
\vphantom{\int}
	\times\{-2\im[D_{0\bar\sigma\bar\sigma0}^r(\omega)
		+\eta_{\bar\sigma}D_{\sigma2\bar\sigma0}^r(\omega)]\}d\omega,
\label{eq-loop1}
\end{eqnarray}
\begin{eqnarray}
\delta\Delta_{2\bar\sigma}=-\frac{\eta_{\bar\sigma}}{2\pi}
	\sum_{\alpha=L,R}\sum_{k\in\alpha}|v_{k\bar\sigma}|^2
	\int\frac{f_\alpha(\dote{k\bar\sigma})-f(\omega)}
			{\dote{k\bar\sigma}-\omega}
\nonumber\\
\vphantom{\int}
	\times\{-2\im[D_{0\bar\sigma2\sigma}^r(\omega)
		+\eta_{\bar\sigma}D_{\sigma22\sigma}^r(\omega)]\}d\omega,
\label{eq-loop2}
\end{eqnarray}
\end{subequations}
where $D_{a\bar{b}}^r(\omega)$ is the retarded form of the corresponding locator, and $f_{L/R}(\omega)=f(\omega-\mu_{L/R})$ is the Fermi function at the chemical potential $\mu_{L/R}$ for the left/right lead. The renormalisation due to the loop correction and its effects on the transport properties is discussed elsewhere \cite{franssonPRL2002,franssonPRB2004,franssonPRL2005}.

The formal structure of the equation for the QD GF, e.g. Dyson equation, gives by a direct application of the Langreth rules for analytical continuation \cite{langreth1976} the retarded/advanced and lesser counterparts of the GF as
\begin{equation}
\begin{array}{rcl}
\bfG^{r/a}(\omega)&=&\bfg^{r/a}(\omega)
	+\bfg^{r/a}(\omega)\bfV^{r/a}(\omega)\bfG^{r/a}(\omega),
\\
\bfG^<(\omega)&=&\bfG^r(\omega)\bfV^<(\omega)\bfG^a(\omega).
\end{array}
\label{eq-Gral}
\end{equation}
Here $\bfg$ is the GF for the QD in the atomic limit, whereas the components of $\bfV^<(\omega)$ are given by $V_\sigma^<(\omega)=i[f_L(\omega)\Gamma_\sigma^L+f_R(\omega)\Gamma_\sigma^R]$; $\Gamma_\sigma=\Gamma^L_\sigma+\Gamma^R_\sigma$ and $\Gamma^{L/R}_\sigma=-2\im\sum_{k\in L/R}|v_{k\sigma}|^2g_{k\sigma}^r(\omega)$. The retarded/advanced and lesser GFs can now be inserted into the general expression for the current, Eq. (19) in Ref. \cite{jauho1994}, where the trace is taken over the $4\times4$ matrices. It is then straightforward to show that
\begin{subequations}
\label{eq-trace}
\begin{eqnarray}
\lefteqn{
\tr[f_L(\omega)\bfGamma^L-f_R(\omega)\bfGamma^R]
	[\bfG^r(\omega)-\bfG^a(\omega)]=
}
\nonumber\\&&\hspace{-0.3cm}
	=-i\sum_\sigma[f_L(\omega)\Gamma_\sigma^L-f_R(\omega)\Gamma_\sigma^R]
		\Gamma_\sigma|{\cal G}_\sigma^r(\omega)|^2,
\label{eq-traceGra}
\\
\lefteqn{
\tr[\bfGamma^L-\bfGamma^R]\bfG^<(\omega)=
}
\nonumber\\&&\hspace{-0.3cm}
	=i\sum_\sigma[\Gamma_\sigma^L-\Gamma_\sigma^R]
		[f_L(\omega)\Gamma_\sigma^L+f_R(\omega)\Gamma_\sigma^R]
		|{\cal G}_\sigma^r(\omega)|^2,
\label{eq-traceGl}
\end{eqnarray}
\end{subequations}
with ${\cal G}_\sigma^r=G_{0\sigma\sigma0}^r+\eta_{\sigma}[G_{0\sigma2\bar\sigma}^r+G_{\bar\sigma2\sigma0}^r]+G_{\bar\sigma22\bar\sigma}^r$, giving
\begin{eqnarray}
{\cal G}_\sigma^r(\omega)&=&
	(\omega-\Delta_{\sigma0}^0-\delta\Delta_{\sigma0}-\delta\Delta_{2\bar\sigma}
		-UP_{0\sigma})
\nonumber\\&&
	\times[(\omega-\Delta_{\sigma0}^0-V_\sigma^r)
	(\omega-\Delta_{2\bar\sigma}^0
		-\delta\Delta_{\sigma0}-\delta\Delta_{2\bar\sigma})
\nonumber\\&&
	-U(P_{\bar\sigma2}V_\sigma^r-\delta\Delta_{\sigma0})]^{-1},
\label{eq-G}
\end{eqnarray}
where $P_{0\sigma}=P_{0\sigma\sigma0}$ and $P_{\bar\sigma2}=P_{\bar\sigma22\bar\sigma}$, and $V_\sigma^r=V_\sigma^r(\omega)$. Hence, the current becomes
\begin{equation}
J=\frac{e}{h}\sum_\sigma\int\Gamma_\sigma^L\Gamma_\sigma^R
	|{\cal G}_\sigma^r(\omega)|^2
	[f_L(\omega)-f_R(\omega)]d\omega.
\label{eq-J}
\end{equation}
It should be emphasised that the derivation of Eq. (\ref{eq-J}) is valid for arbitrary couplings $\bfGamma^{L/R}$, \emph{not} only $\bfGamma^L=\lambda\bfGamma^R$.

Here, the end-factors are given by $P_{0\sigma}=\im\int[G_{0\sigma}^<(\omega)-G_{\sigma2}^>(\omega)-\sum_{\sigma'}G_{0\sigma'}^>(\omega)]d\omega/(2\pi)$ and $P_{\bar\sigma2}=\im\int[G_{0\bar\sigma}^<(\omega)-G_{\bar\sigma2}^>(\omega)+\sum_{\sigma'}G_{\sigma'2}^<(\omega)]d\omega/(2\pi)$, which leads to the boundary condition $P_{0\sigma}+P_{\bar\sigma2}=-\tr\im\int\bfG^r(\omega)d\omega/\pi=1$ \cite{Gl-Gr} by the requirement that the integrated total local density of states is unity \cite{sandalov2003,franssonPRL2005}. Hence, Eqs. (\ref{eq-loop}) and (\ref{eq-Gral}), along with of the end-factors, lead to a set of equations subject to self-consistent calculations, for each value of the bias voltage. This, then provide an accurate field dependence of the transport properties of the system. Replacing $\bfGamma^L-\bfGamma^R$ by $\bfGamma$ and $f_L\bfGamma^L-f_R\bfGamma^R$ by $f_L\bfGamma^L+f_R\bfGamma^R$ in Eq. (\ref{eq-trace}), one easily shows charge and current conservation of the presented result, e.g. $\ddtinline(P_{0\sigma}+P_{\bar\sigma2})=0$.

Now, since the occupation number $\av{n_\sigma}=\av{\h{\sigma}{}+\h{2}{}}=P_{\sigma2}$, it follows that $P_{0\sigma}=1-\av{n_{\bar\sigma}}$. Hence, removing the renormalisation energies in Eq. (\ref{eq-G}), reduces the GF to the standard mean-field result, e.g. the HIA \cite{hewson1966}.

The atomic limit of Eq. (\ref{eq-G}) is trivial since this limit is the starting point of the expansion of the QD GF. However, it is  worth to notice that $\delta\Delta_{\sigma0},\delta\Delta_{2\bar\sigma}\rightarrow0$ as $v_{k\sigma}\rightarrow0$, c.f. Eq. (\ref{eq-loop}). Further, in the limit of no on-site Coulomb interaction, e.g. $U\rightarrow0$, the renormalisation energies $\delta\Delta_{\sigma0}+\delta\Delta_{2\bar\sigma}\rightarrow0$ \cite{franssonPRB2005}, giving ${\cal G}_\sigma^r(\omega)=(\omega-\dote{0}-V_\sigma^r)^{-1}$, i.e. the result from the exactly solvable Fano-Anderson model \cite{fano1961,anderson1961}. Inserting this expression into the current, Eq. (\ref{eq-J}), yields the well-known result for the single resonant level \cite{larkin1987,jauho1994}. Finally, in the limit of strong on-site Coulomb repulsion, e.g. $U\rightarrow\infty$, Eq. (\ref{eq-G}) becomes ${\cal G}_\sigma^r(\omega)=P_{0\sigma}/(\omega-\Delta_{\sigma0}-P_{0\sigma}V_\sigma^r)$, where $\Delta_{\sigma0}=\Delta_{\sigma0}^0+\delta\Delta_{\sigma0}$ and (at $T=0$ K)
\[ \av{n_\sigma}=\frac{P_{0\sigma}}{\Gamma_\sigma}
	\sum_{\alpha=L,R}\Gamma_\sigma^\alpha\biggl[
		\frac{1}{\pi}
			\arctan{\frac{\mu_\alpha-\Delta_{\sigma0}}
				{P_{0\sigma}\Gamma_\sigma/2}}
			+\frac{1}{2}\biggr],
\]
since $\delta\Delta_{\sigma0},\delta\Delta_{2\bar\sigma}$ are finite for all $U$. This GF was derived in Ref. \cite{franssonPRL2002} and is consistent with the result in Ref. \cite{barabanov1974}. By removing the loop correction $\delta\Delta_{\sigma0}$, this GF reduces to the result in Ref. \cite{varma1976}. Also in this case the current becomes particularly simple, e.g. 
\[ J=\frac{e}{h}\sum_\sigma\int
	\frac{\Gamma_\sigma^L\Gamma_\sigma^RP_{0\sigma}^2}
	{|\omega-\Delta_{\sigma0}-P_{0\sigma}V_\sigma^r|^2}
	[f_L(\omega)-f_R(\omega)]d\omega.
\]
The difference between this expression and the one for the single resonant level is that the end-factor $P_{0\sigma}$ makes sure that the localised level is at most occupied by one electron, and that the transition energy is appropriately renormalised due to the strong electron correlations.

In the remainder of this Letter, the difference in the transport properties of the HIA and the loop correction in the non-magnetic case, is addressed. Therefore, let $\Gamma_\sigma=\Gamma$, and let $\mu=0$ be the equilibrium chemical potential. In general, the renormalisation of the transition energies tends to increase the effective Coulomb repulsion, e.g.  $\tilde{U}=\Delta_{2\bar\sigma}^0+\delta\Delta_{2\bar\sigma}-\Delta_{\sigma0}^0-\delta\Delta_{\sigma0}>U$ since $\Delta_{2\bar\sigma}^0-\Delta_{\sigma0}^0=U$ and $\delta\Delta_{\sigma0}\leq0\leq\delta\Delta_{2\bar\sigma}$. However, for $U/\Gamma\ll1$ and $U/\Gamma\gg1$, this does not cause any dramatic deviation between the two approximations in the resulting transport properties.

\begin{figure}[t]
\begin{center}
\includegraphics[width=8cm]{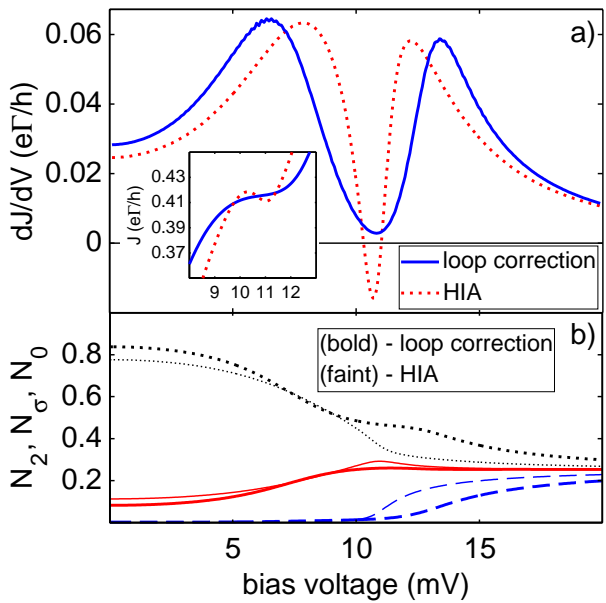}
\end{center}
\caption{(Colour online). Calculated transport characteristics of the single level QD. a) Differential conductance calculated in the HIA (dotted) and with the loop correction (solid) (the small ripples in the solid line is due to that $dJ/dV$ is the numerical derivative of the calculated current). Inset: The current in the voltage interval between the conductance peaks. b) Occupation numbers $N_0$ (dotted), $N_\sigma\ (N_\up=N_\down)$ (solid), and $N_2$ (dashed) in the HIA (faint) and with the loop correction (bold). Here, $\{\dote{0},U,k_BT\}/\Gamma=\{1,0.5,0.02\}$.}
\label{fig-jv}
\end{figure}

In contrast, by tuning the system into the regime $1<\Delta_{\sigma0}^0/\Gamma<\Delta_{2\bar\sigma}^0/\Gamma$ (or $\Delta_{\sigma0}^0/\Gamma<\Delta_{2\bar\sigma}^0/\Gamma<-1$) and $0.5\lesssim U/\Gamma\lesssim1$, for low temperatures, one finds that the current resulting from the HIA is weakly peaked for bias voltages such that either of the chemical potentials $\mu_{L/R}$ lies between the two transition energies, see inset of Fig. \ref{fig-jv} a) (dotted). Consequently, there will be a region of a clear negative differential conductance (NDC) between the two conductance peaks corresponding to the transition energies $\Delta_{\sigma0}$ and $\Delta_{2\bar\sigma}$, Fig. \ref{fig-jv} a) (dotted). This resonant peak is, however, not consistent with recent experimental data on single-level QDs, for example see Ref. \cite{exp}. In the approximation with the loop correction (solid), on the other hand, the differential conductance is positive in the whole bias voltage interval, in agreement with experiments. The current then shows a plateau, as expected when the one-particle state is resonant while the two-particle state is not.

The difference of the two solutions can be understood as follows. The renormalisation of the transition energies yields $\Delta_{\sigma0}\leq\Delta_{\sigma0}^0$  and $\Delta_{2\bar\sigma}\geq\Delta_{2\bar\sigma}^0$. Hence, the state $\ket{\sigma}$ begins to populate at lower bias voltages, Fig. \ref{fig-jv} b) (bold solid), than in the HIA, whereas the state $\ket{2}$ remains unoccupied for higher voltages (bold dashed). Therefore, the population in the one-particle state is able to saturate for bias voltages such that the two-particle state remains unoccupied. In the HIA, on the other hand, the one- and two-particle states compete about the available population, since the transition energies lie closer to one another, which leads to an overpopulation of the one-particle state, Fig. \ref{fig-jv} b) (faint solid). In general, a high population number in the one-particle state in combination with a significant reduction in the population number of the empty state $\ket{0}$, tends to reduce the tunnelling probability of the one-particle state \cite{franssonPRB2004}. Hence, the overpopulation in $\ket{\sigma}$ affects the transport properties of the system such that the current decreases. For larger bias voltages, however, the population in the one-particle state returns to its normal saturation value leading to an increasing current.

In conclusion, an analytical formula for the current through a single level QD for arbitrary on-site Coulomb repulsion and couplings to the leads, was derived beyond standard mean-field theory (HIA) using a diagrammatic expansion of the QD GF in the  strong coupling regime. The result removes the region of NDC, found in the HIA, in agreement with recent experimental data.

Support from Carl Trygger Foundation, G\"oran Gustafsson's Foundation, and Swedish Foundation of Strategic Research (SSF) is acknowledged.


\begin{thebibliography}{20}
\bibitem{kondo} D. Goldhaber-Gordon, \etal, Nature, {\bf 391}, 156 (1998), S. M. Cronenwett, T. H. Oosterkamp, and L. P. Kouwenhoven, Science, {\bf 281}, 540 (1998); Y. Ji, M. Heiblum, D. Sprinzak, D. Mahalu, and H. Shtrikman, Science, {\bf 290}, 779 (2000).
\bibitem{resonant} R. A. Webb, A. Hartstein, J. J. Wainer, and A. B. Fowler, Phys. Rev. Lett. {\bf 54}, 1577 (1985); A. T. Johnson \etal, \emph{ibid} {\bf 69}, 1592 (1992).
\bibitem{orthodox} D. V. Averin and K. K. Likharev, in \emph{Mesoscopic Phenomena in Solids}, eds. B. L. Altshuler \etal (Elsevier, Amsterdam, 1991).
\bibitem{coherent} J. K\"onig and Y. Gefen, Phys. Rev. Lett. {\bf 86}, 3855 (2001); R. Aguado and D. C. Langreth, \emph{ibid} {\bf 85}, 1946 (2000); T. Brandes and F. Renzoni, \emph{ibid} {\bf 85}, 4148 (2000); L. E. F. Foa Torres, C. H. Lewenkopf, and H. M. Pastawski, \emph{ibid} {\bf 91}, 116801 (2003); D. A. Abanin and L. S. Levitov, \emph{ibid} {\bf 93}, 126802 (2004).
\bibitem{hubbard} J. Hubbard, Proc. Roy, Soc., Ser. A {\bf 276}, 238 (1963); {\bf 277}, 237 (1963).
\bibitem{hewson1966} A. C. Hewson, Phys. Rev. {\bf 144}, 420 (1966).
\bibitem{varma1976} C. M. Varma and Y. Yafet, Phys. Rev. B, {\bf 13}, 2950 (1976).
\bibitem{barabanov1974} A. F. Barabanov, K. A. Kikoin, and L. A. Maksimov, Solid State Comm. {\bf 15}, 977 (1974); {\bf 18}, 1527 (1976).
\bibitem{renormalisation} L. D. Faddeev, Sov. Phys. JETP, {\bf 12}, 1014 (1961); F. D. M. Haldane, Phys. Rev. Lett. {\bf 40}, 416 (1978); A. E. Ruckenstein and S. Schmitt-Rink, Int. J. Mod. Phys. B, {\bf 3} 1809 (1989); Y. A. Izyumov, B. M. Letfulov, E. V. Shipitsyn, and K. A. Chao, Int. J. Mod. Phys. B, {\bf 6}, 3479 (1992); Y. A. Izyumov, B. M. Letfulov, and E. V. Shiptisyn, J. Phys. Condens. Matter, {\bf 6}, 5137 (1994).
\bibitem{anderson1961} P. W. Anderson, Phys. Rev. {\bf 124}, 41 (1961).
\bibitem{sandalov2003} I. Sandalov, B. Johansson, and O. Eriksson, Int. J. Quantum Chem. {\bf 94}, 113 (2003).
\bibitem{franssonPRL2002} J. Fransson, O. Eriksson, and I. Sandalov, Phys. Rev. Lett. {\bf 88}, 226601 (2002); {\bf 89}, 179903 (2002).
\bibitem{franssonPRB2004} J. Fransson, Phys. Rev. B, {\bf 69}, 201304(R), (2004); J. Fransson and O. Eriksson, \emph{ibid} {\bf 70}, 085301 (2004).
\bibitem{franssonPRL2005} J. Fransson, \emph{submitted} (2004).
\bibitem{jauho1994} A. -P. Jauho, N. S. Wingreen, and Y. Meir, Phys. Rev. B, {\bf 50}, 5528 (1994).
\bibitem{kadanoff} L. P. Kadanoff and G. Baym, \emph{Quantum Statistical Mechanics} (Benjamin, New York, 1962).
\bibitem{langreth1976} D. C. Langreth, in \emph{Linear and Nonlinear Electron Transport in Solids}, Vol. 17 of \emph{NATO Advanced Study Institute, Series B: Physics}, eds. J. T. Devreese and V. E. van Doren (Plenum, New York, 1976).
\bibitem{Gl-Gr} Using the identity $\bfG^>-\bfG^<=\bfG^r-\bfG^a$.
\bibitem{franssonPRB2005} J. Fransson, \emph{in preparation} (2005).
\bibitem{fano1961} U. Fano, Phys. Rev. Lett. {\bf 124}, 1866 (1961).
\bibitem{larkin1987} A. I. Larkin and K. A. Matveev, Sov. Phys. JETP, {\bf 66}, 580 (1987).
\bibitem{exp} S. H. Persson, L. Olofsson, and L. Gunnarsson, Appl. Phys. Lett. {\bf 74}, 2546 (1999); M. Jung \etal, Appl. Phys. Lett. {\bf 86}, 033106 (2005).
\end{thebibliography}
\end{document}